\documentclass[prd,preprint,superscriptaddress,preprintnumbers,eqsecnum,showpacs,nofootinbib,nobibnotes]{revtex4}
\usepackage{amsfonts,bm}
\usepackage{graphicx}
\usepackage{pstricks}

\newcommand{\be}{\begin{equation}}
\newcommand{\bea}{\begin{eqnarray}}
\newcommand{\ee}{\end{equation}}
\newcommand{\eea}{\end{eqnarray}}
\newcommand{\bpi}{\begin{picture}}
\newcommand{\bce}{\begin{center}}
\newcommand{\epi}{\end{picture}}
\newcommand{\ece}{\end{center}}

\def\gtree{\Gamma^{(0)}}
\def\gtreeb{\widetilde{\Gamma}^{(0)}}
\def\gfullb{\widetilde{\Gamma}}
\def\bqq{\widetilde{\mathrm{I}\!\Gamma}}

\def\bcj{J}
\def\bcjb{\widetilde{\bcj}}

\def\dim{d}

\def\diff{{\rm d}}

\begin{document}

\title{Gauge invariant Ansatz for a special three-gluon vertex}
\date{February 28, 2011}

\author{D. Binosi}
\email{binosi@ect.it}
\affiliation{European Centre for Theoretical Studies in Nuclear
Physics and Related Areas (ECT*) and Fondazione Bruno Kessler, \\Villa Tambosi, Strada delle
Tabarelle 286, 
I-38123 Villazzano (TN)  Italy}

\author{J. Papavassiliou}
\email{Joannis.Papavassiliou@uv.es}
\affiliation{\mbox{Department of Theoretical Physics and IFIC,  
University of Valencia}
E-46100, Valencia, Spain}

\begin{abstract}
We  construct  a  general  Ansatz  for  the
three-particle vertex describing the interaction of one background and
two   quantum  gluons,   by  simultaneously   solving  the   Ward  and
Slavnov-Taylor identities  it satisfies.  This  vertex is known  to be
essential  for the gauge-invariant truncation of the  
Schwinger-Dyson   equations  of QCD,   
based  on  the pinch technique and the background field  method. 
A key step in this construction is  the formal derivation of a set of
crucial constraints  (shown to  be valid to  all orders),  relating the
various form  factors of the  ghost Green's functions 
appearing  in the aforementioned Slavnov-Taylor identity. 
When inserted into the Schwinger-Dyson equation for the gluon propagator, 
this vertex gives rise to a number of highly non-trivial cancellations, 
which are absolutely indispensable for the self-consistency of the 
entire approach.

\end{abstract}


\maketitle

\section{Introduction}

\noindent 

\noindent In recent years, a significant part of the ongoing activity dedicated to
the  study the  non-perturbative  sector of  Yang-Mills theories,  and
especially of QCD, has focused on the infrared behavior of individual
Green's  functions.    The  information  obtained  by   a  variety  of
large-volume lattice  simulations has been of  singular importance for
advancing  in   this  direction,   and  has  stimulated   an  in-depth
re-examination of various aspects  of the underlying QCD dynamics.  In
particular,  these lattice  results  clearly indicate  that the  gluon
propagator  and  the  ghost   dressing  function  of  pure  Yang-Mills
theories,  computed in  the  conventional Landau  gauge, are  infrared
finite,    both     in    $SU(2)$~\cite{Cucchieri:2007md,Cucchieri:2007rg,Cucchieri:2009zt,Cucchieri:2011ga,Cucchieri:2011um}    and    in
$SU(3)$~\cite{Bogolubsky:2007ud,Bowman:2007du,Bogolubsky:2009dc,Oliveira:2009eh}.

Evidently,  reproducing  these  (and  related) lattice  results  using
continuous approaches  represents a highly  non-trivial challenge.  In
this  effort,  the  Schwinger-Dyson  equations  (SDEs)  constitute  an
obvious (albeit  technically cumbersome)  starting point. As  has been
argued          in         a         series          of         recent
articles~\cite{Aguilar:2006gr,Binosi:2007pi,Binosi:2008qk}, the modified set of SDEs
obtained  within the general  formalism based  on the  Pinch Technique
(PT)~\cite{Cornwall:1981zr,Cornwall:1989gv,Binosi:2002ft,Binosi:2003rr,Binosi:2009qm}  and  the  Background  Field
Method  (BFM)~\cite{Abbott:1980hw},  is  particularly well-suited  for
attempting this difficult task (for alternative approaches see, {\it e.g.,}~\cite{Alkofer:2000wg,Fischer:2006ub,Dudal:2008sp,Boucaud:2008ji,Braun:2007bx,Szczepaniak:2010fe,RodriguezQuintero:2010ss,RodriguezQuintero:2010yq}).

One of the key ingredients of the PT-BFM approach, in general, is (see Fig.~\ref{3g-vertex})
the three-gluon vertex  describing the interaction 
between one background ($B$)  and two quantum ($Q$) gluons 
(``$BQQ$ vertex'', for short). 
This vertex appears naturally in the modified SDE governing the gluon self-energy,  
and is instrumental for its gauge-invariant truncation. 
In particular, and contrary to what happens in the conventional formulation, 
the ``one-loop dressed'' subset 
of (only gluonic!) diagrams (see Fig.~\ref{gSDE}), 
corresponding to the first step in the aforementioned SDE truncation,
furnishes an exactly transverse  gluon self-energy.
In addition, the way the gluon acquires 
a dynamically generated (momentum-dependent) mass~\cite{Cornwall:1981zr}, 
which, in turn, accounts for the infrared  finiteness of the aforementioned 
Green's functions, is determined by a subtle interplay 
of all the ingredients entering into the corresponding SDE.
In this context  the non-perturbative form of the $BQQ$ vertex is 
essential for obtaining infrared finite results out of the SDEs considered, 
without violating any of the underlying field-theoretic principles~\cite{Aguilar:2009ke}. 

A major difficulty that is typical in the SDE studies 
(and not only in the case of the $BQQ$ vertex considered here) 
is precisely the 
form that one must use for the various fully-dressed vertices 
entering into the problem.
To be sure, the non-perturbative behavior of each such vertex (including the $BQQ$ vertex) 
is determined by its
own SDE equation, which contains the various multiparticle kernels appearing 
in the ``skeleton expansion'' (see Fig.~\ref{SDE-vertex}). 
However, for practical purposes, one is 
forced to resort to an Ansatz for this vertex, obtained through the so-called 
``gauge-technique''~\cite{Salam:1963sa,Salam:1964zk,Delbourgo:1977jc,Delbourgo:1977hq}.

The idea behind the gauge-technique is fairly simple, even though its precise implementation 
may be rather complicated.  
Specifically, one constructs an expression for the unknown 
vertex out of the ingredients appearing in the Ward identity (WI) and/or the Slavnov-Taylor identity (STI) it satisfies. 
These ingredients must be put together in a way such that 
the resulting expression satisfies the WI and/or the STI automatically.
Evidently, this technique becomes more difficult to implement as the Lorentz and color structure of the vertex under construction increases, and the structure of the STIs that it satisfies gets more involved. In addition,  
it is clear that this method can only determine the 
``longitudinal'' part of any vertex, leaving its 
``transverse'' ({\it i.e.}, automatically conserved) part completely unspecified.
Failing  to provide  the correct
transverse  part  leads  to  the mishandling of  overlapping  divergences,
which,  in turn, compromises  the multiplicative  renormalizability of
the resulting SDE. The usual remedy employed in the literature 
is to account approximately for the missing pieces by modifying appropriately (but``by hand'') 
the SDE in question. 

In this article we will cary out in detail the gauge-technique construction for the  $BQQ$ vertex 
mentioned above. This is a particularly involved task, and deviates 
appreciably from the corresponding construction of the 
conventional three-gluon vertex (the ``$QQQ$ vertex'' in this notation)~\cite{Ball:1980ax}, 
mainly for the following reasons.

\begin{itemize}
\item[({\it i})] Unlike the $QQQ$ vertex, which displays Bose symmetry with respect to the 
interchange of any one of its three legs, the $BQQ$ vertex is Bose symmetric only 
under the interchange of the two quantum legs. As a result, the constraints imposed 
by Bose symmetry on the various form-factors comprising the two vertices are 
different.  

\item[({\it ii})] Whereas the $QQQ$ vertex satisfies the same STI 
when contracted from any direction,
the   $BQQ$ vertex satisfies an Abelian (ghost-free) WI when contracted from the 
side of the background gluon, and an STI when contracted from the side of 
either one of the quantum gluons [see Eq.~(\ref{sward})].

\item[({\it iii})] The number of Green's functions and composite (BRST-induced) 
operators entering  
into the  STI satisfied by the $BQQ$ is practically duplicated 
compared to the $QQQ$ case. 
Indeed, while the 
the STI maintains its basic characteristic form, any Green's function 
that appears in it and has an incoming gluon ({\it i.e.}, gluon self-energy or the 
``gluon-ghost'' kernel) appears in two versions: in the first, the incoming gluon is a quantum gluon, 
in the second, it is a background one [the latter quantities are denoted by ``tildes''
on the rhs of Eq.~(\ref{sward})].
\end{itemize}

As is well-known from the case of the conventional vertex~\cite{Ball:1980ax}, 
the gauge-technique construction boils down finally to the solution of a system of various equations, 
whose unknowns are the form factors (of the longitudinal part) of the vertex under construction.
The solution of this system allows to express these form factors in terms 
of the various quantities appearing on the rhs of the STI:
gluon propagator(s), ghost dressing function, and a subset of the form factors of the ``gluon-ghost'' 
kernel(s). However, solving the resulting system is conceptually subtle, 
because the additional constrains imposed by Bose-symmetry reduces the number of 
unknowns, and one is left with more equations than unknowns. Thus, in order to 
find non-trivial solutions, a set of additional identities must be imposed, 
which relates the ingredients appearing on the rhs of the STI; 
in particular, the ghost dressing function is related to some of the 
form-factors of the ``gluon-ghost'' kernels [see Eqs.~(\ref{Hhatconstr}) and~(\ref{Hconstr})].
This reduces the number 
of available equations, because some of them are identically satisfied, precisely 
by virtue of these additional identities. These identities can be established 
by inspection, {\it i.e.}, as a necessary condition for having 
non-trivial solutions for the system. However, given that they 
constitute, at the same time, non-trivial relations between well-defined field-theoretic 
quantities (those appearing on the rhs of the STI), their validity must  
hold regardless of the need to solve the given system of equations.     
In the work of~\cite{Ball:1980ax} the aforementioned set of crucial auxiliary identities 
has been  established as a necessary condition for solving the system, and their 
validity has been indeed confirmed  at the one-loop perturbative level, through 
an explicit calculation (the complete one-loop off-shell form factors, in an arbitrary covariant gauge $\xi$ and space-time dimension, 
have been calculated in~\cite{Davydychev:1996pb}).

In the case of the $BQQ$ vertex we consider, and given the aforementioned 
duplication of the quantities entering into the STI, 
the solution of the system requires the validity of two types of 
such auxiliary identities: one of them coincides with that found in~\cite{Ball:1980ax},  
while the other is completely new, and reported here for the first time. 
It turns out that the validity of both identities can be demonstrated 
to all-orders (and non-perturbatively); indeed, they are a direct consequence of  
a STI and a WI that the two ghost sectors (the ``conventional'' one and the ``tilded'' one, respectively) satisfy. 
To the best of our knowledge this is a novel result.     

The article is organized as follows. 
In Section~\ref{BQQprop} we present some basic facts about the $BQQ$ vertex. We pay particular attention to the WI and STI this vertex satisfies, and explain in detail 
the definition and field-theoretic origin of the various quantities entering in them.   
In Section~\ref{ghost}
we resort to the Batalin-Vilkovisky formalism,  
in order to derive the auxiliary STI and WI satisfied by the 
two types of ghost-induced Green's functions appearing in the central STI 
(satisfied by $BQQ$). These two auxiliary equations 
are valid to all orders, and 
give rise to the set of constraints needed for the solution of the system in the next section. 
Section~\ref{syst} contains the main result of this article. 
Specifically, the system of equations involving 
the form factors of the longitudinal part of the 
vertex is presented, and its solution is reported, after using  
the additional constraints derived in the previous section. 
In Section~\ref{cons} 
we give a detailed account of the most important theoretical consequences 
that the precise form of the 
vertex has for the SDE of the gluon propagator (in the Landau gauge). 
Finally, in Section~\ref{concl} we present our conclusions.

\section{The $BQQ$ vertex and its basic properties}
\label{BQQprop}

\noindent The $BQQ$ vertex constitutes without any doubt one of the most fundamental ingredients 
of the pinch technique, making its appearance already at the basic level of the one-loop construction. 
Specifically,  defining the tree-level conventional 
three-gluon vertex through the expression (all momenta entering)
\bea
i\Gamma^{(0)}_{A^a_\alpha A^b_\mu A^c_\nu}(q,r,p)&=&gf^{abc}\gtree_{\alpha\mu\nu}(q,r,p)\nonumber \\
\gtree_{\alpha\mu\nu}(q,r,p)&=&g_{\mu\nu}(r-p)_\alpha  +
g_{\alpha\nu}(p-q)_\mu+g_{\alpha\mu}(q-r)_\nu,
\eea
the diagrammatic rearrangements giving rise to the PT Green's functions (propagators and vertices) stem exclusively from the  characteristic decomposition~\cite{Cornwall:1981zr,Cornwall:1989gv,Pilaftsis:1996fh}
\begin{eqnarray}
\gtree_{\alpha\mu\nu}(q,r,p)&=& \gtreeb_{\alpha\mu\nu}(q,r,p)+ (1/\xi)
\Gamma^{{\rm P}}_{\alpha\mu\nu}(q,r,p),     \nonumber  \\ 
\gtreeb_{\alpha\mu\nu}(q,r,p)&=&g_{\mu\nu}(r-p)_\alpha  +
g_{\alpha\nu}(p-q+ r/\xi)_\mu+g_{\alpha\mu}(q-r- p/\xi)_\nu, \nonumber \\    
\Gamma^{{\rm P}}_{\alpha\mu\nu}(q,r,p)&=& g_{\alpha\mu} p_\nu - g_{\alpha\nu}r_\mu.
\label{deco}
\end{eqnarray}
In the equations above, $\xi$ represents the gauge-fixing parameter that  appears also in the definition of the (full) gluon propagator $\Delta^{ab}_{\mu\nu}(q)=\delta^{ab}\Delta_{\mu\nu}(q)$, with
\be
i\Delta_{\mu\nu}(q)=- i\left[P_{\mu\nu}(q)\Delta(q^2)+\xi\frac{q_\mu q_\nu}{q^4}\right]; \qquad
\Delta^{-1}_{\mu\nu}(q)=i\left[P_{\mu\nu}(q)\Delta^{-1}(q^2)+(1/\xi) q_\mu q_\nu\right]
\label{prop}
\ee
and $P_{\mu\nu}(q)=g_{\mu\nu}-q_\mu q_\nu/q^2$ the dimensionless transverse projector; finally, the scalar cofactor $\Delta(q^2)$ is finally related to the all-order gluon self-energy $\Pi_{\mu\nu}(q)=P_{\mu\nu}(q)\Pi(q^2)$  through
\be
\Delta^{-1}({q^2})=q^2+i\Pi(q^2)=q^2\bcj(q^2).
\ee

Notice that the PT makes no {\it ab initio}  
reference to a background gluon; at the level of the Yang-Mills Lagrangian there is only one gauge field, $A$, 
which is quantized in the usual way, by means of a linear gauge-fixing function of the $R_\xi$ type ${\cal F}^a=\partial^{\mu} A^a_{\mu}$. 
However, the decomposition~(\ref{deco}) assigns 
right from the start a special role to the leg carrying the momentum $q$, that is to be eventually identified with the background leg.  
Thus, unlike $\gtree$, which is Bose-symmetric with respect to all its three legs,  
the vertex $\gtreeb$ is in fact Bose-symmetric {\it only} with respect to the (quantum) $\mu$ and $\nu$ legs. In addition, 
it satisfies the simple Ward identity
\be
iq^\alpha \gtreeb_{\alpha\mu\nu}(q,r,p)=\Delta^{-1}_{0\,\mu\nu}(p)- \Delta^{-1}_{0\,\mu\nu}(r),
\ee
where the sub-index ``0'' on the rhs
indicates the tree-level version of the inverse propagator (\ref{prop}).
In higher orders, the $BQQ$ vertex is constructed  
through the systematic triggering of internal STIs in the diagrams of the conventional 
(higher order) three-gluon vertex~\cite{Binosi:2002ft,Binosi:2003rr}. 

\begin{figure}[!t]
\includegraphics[scale=.65]{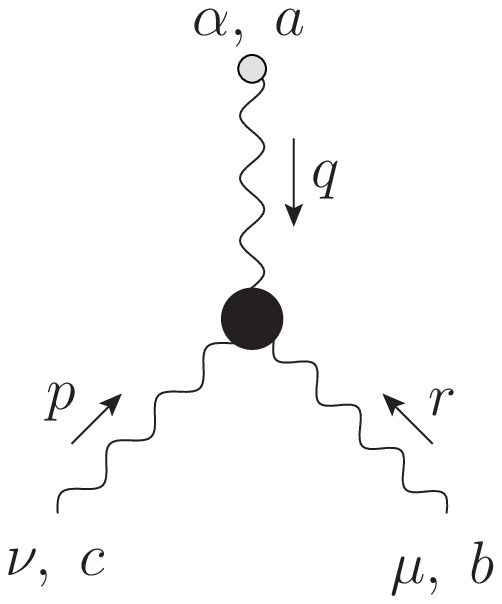}
\caption{\label{3g-vertex}The $BQQ$ three-gluon vertex. The background leg is indicated by the gray circle.}
\end{figure}

On the other hand, when quantizing the theory within the BFM, the notion of a 
$BQQ$ vertex $\Gamma_{\widehat{A}AA}$ arises naturally as a consequence of the 
splitting of the classical gauge field into a background and a quantum part, $A\to A+\widehat{A}$, 
and the choice of a special gauge fixing function ${\cal F}^a=\partial^{\mu} A^a_{\mu}+gf^{abc}\widehat{A}^{\mu}_b A_\mu^c$,
which is linear in the quantum field $A$, and  preserves gauge invariance with respect to the background field $\widehat{A}$. 
The latter induces to the tree-level $BQQ$ vertex an additional dependence on the gauge-fixing parameter $\xi_Q$, 
and one has~(see Fig.~\ref{3g-vertex})
\be
i\Gamma_{\widehat{A}^a_\alpha A^b_\mu A^c_\nu}(q,r,p)=gf^{abc}\gfullb_{\alpha\mu\nu}(q,r,p)\label{bqq-1};
\ee
the tree-level value of the vertex in Eq.~($\ref{bqq-1}$), namely 
$\gtreeb$, coincides with the PT expression  of Eq.~(\ref{deco}), after the replacement $\xi\to\xi_Q$. 
This equality between the PT and BFM construction appears to be rather general;  
for the usual two- and three-point functions 
(such as the gluon propagator, quark-gluon vertex, and three-gluon vertex) 
it has been shown to hold both perturbatively (to all orders)~\cite{Binosi:2002ft,Binosi:2003rr}, 
and non-perturbatively (at the SDE level)~\cite{Binosi:2007pi,Binosi:2008qk}.

\begin{figure}[!t]
\includegraphics[scale=.85]{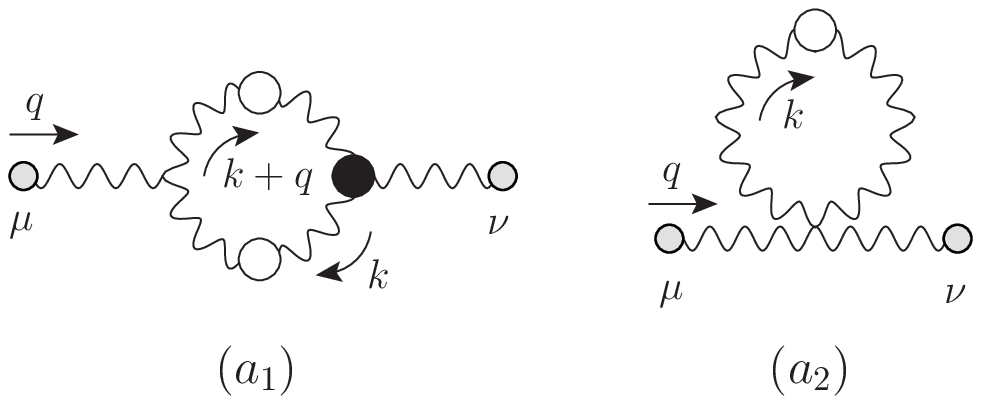}
\caption{\label{gSDE}The one-loop dressed gluon contribution to the  PT-BFM gluon self-energy. 
Notice that, contrary to what happens within the conventional formalism these 
two diagrams constitute a transverse subset of the full SDE.}
\end{figure}

\begin{figure}[!t]
\includegraphics[scale=.65]{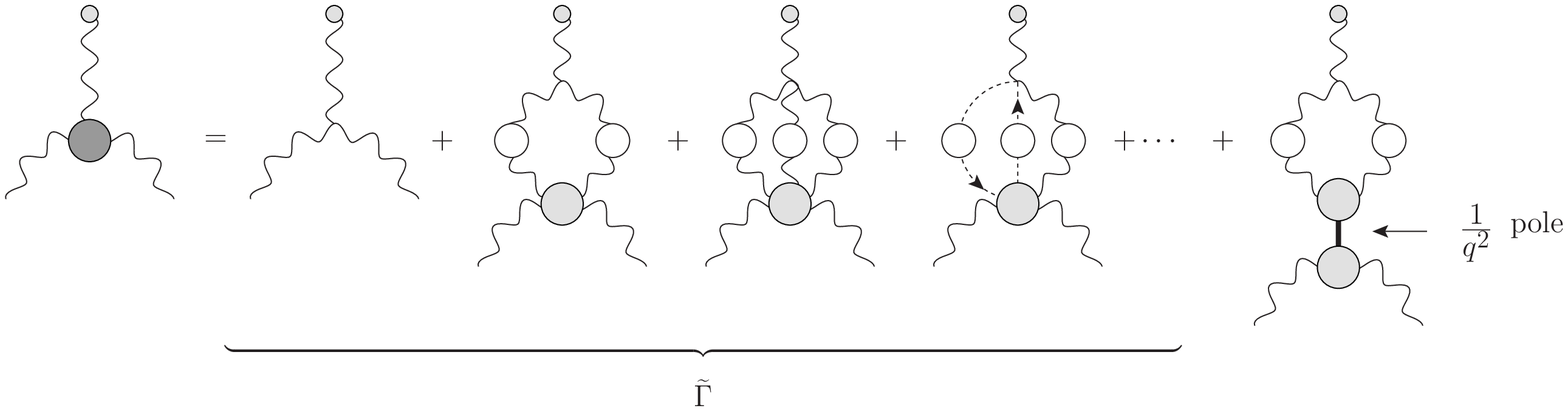}
\caption{\label{SDE-vertex}The SDE equation for the $BQQ$ vertex $\gfullb$. 
The first diagrams correspond to the standard skeleton expansion in terms of  
one-particle irreducible multi-leg kernels (grey blobs).
The last term, representing a non-perturbative bound-state pole,  
appears only when the Schwinger mechanism~\cite{Schwinger:1962tn,Schwinger:1962tp} is triggered~\cite{Jackiw:1973tr,Cornwall:1973ts,Eichten:1974et}.}
\end{figure}

As mentioned in the Introduction, the $BQQ$ vertex $\gfullb$ enters into the SDE 
satisfied by the PT-BFM gluon propagator. Specifically, considering the (gauge-invariant) 
subset of fully dressed diagrams of Fig.~\ref{gSDE},  one finds 
\bea
(a_1)_{\mu\nu}&=& \frac12\,g^2C_A
\int_k\!\gtreeb_{\mu\alpha\beta}\Delta^{\alpha\rho}(k)\Delta^{\beta\sigma}(k+q)\gfullb_{\nu\rho\sigma}\nonumber \\
(a_2)_{\mu\nu}&=&g^2C_A \left[g_{\mu\nu}\int_k\!\Delta^\rho_\rho
+ \left(1/\xi-1\right)\int_k\!\Delta_{\mu\nu}\right],
\label{gl-1ldr}
\eea
with the $\dim$-dimensional integral measure (in dimensional regularization) defined as
\be
\int_{k}\equiv\frac{\mu^{\epsilon}}{(2\pi)^{\dim}}\!\int\!\mathrm{d}^\dim k.
\label{dqd}
\ee
Evidently, non-perturbative information on the vertex $\gfullb$ is essential for making further progress with these equations. 
In principle, the complete structure 
of $\bqq$ is determined from its own SDE; however, this equation is practically intractable, 
given that it involves several unknown one-particle reducible kernels,  
associated with its skeleton expansion shown in Fig.~\ref{SDE-vertex}. 
Given this serious limitation, one usually is forced to approximate the vertex by employing a suitable Ansatz.
In general, such an Ansatz is obtained by resorting to the aforementioned gauge technique~\cite{Salam:1963sa,Salam:1964zk,Delbourgo:1977jc,Delbourgo:1977hq}.  
Even though the actual construction will be carried out in Section~\ref{syst}, it is worthwhile to briefly 
review the basic philosophy behind this technique. 

The main idea is easier captured in the  Abelian context where it was first applied.  
Roughly speaking, one constructs an expression for the unknown 
vertex out of the ingredients appearing in the WI it satisfies. 
These ingredients must be put together in a way such that 
the resulting expression satisfies the WI automatically. 
The most typical example of such a construction is found in the case of the three-particle vertex
of scalar QED, describing the interaction of a photon with a pair of charged scalars.
This vertex, to be denoted by $\Gamma_{\mu}$,  satisfies the abelian all-order WI
\be
q^{\mu}\Gamma_{\mu}= {\mathcal D}^{-1}(k+q) -{\mathcal D}^{-1}(k) \,, 
\label{sward}
\ee
where ${\mathcal D}(k)$  is the fully-dressed propagator of the scalar field.
Thus, in this case,  the gauge-technique Ansatz for $\Gamma_{\mu}$, obtained by Ball and Chiu \cite{Ball:1980ay}, 
after ``solving'' the above WI, under the additional  
requirement of not introducing kinematic singularities, is 
\be
\Gamma_{\mu}= \frac{(2k+q)_{\mu}}{(k+q)^2-k^2}\left[{\mathcal D}^{-1}(k+q) -{\mathcal D}^{-1}(k)\right],
\label{strans_vert}
\ee
which clearly satisfies Eq.~(\ref{sward}). 

Of course, this construction is significantly more complicated for the $BQQ$ vertex $\bqq$, since the identities imposed by the BRST symmetry are far more complex than the simple Abelian WI of Eq.~(\ref{sward}); indeed, in order to cast these upcoming identities into a more compact form, it is convenient to consider, instead of $\gfullb$, the minimally modified vertex $\bqq$, defined as 
\be
\bqq_{\alpha\mu\nu}(q,r,p)=\gfullb_{\alpha\mu\nu}(q,r,p)+(1/\xi_Q)\Gamma^{\rm{P}}_{\alpha\mu\nu}(q,r,p).
\label{bqq}
\ee
Evidently, $\bqq_{\alpha\mu\nu}(q,r,p)$ and $\gfullb_{\alpha\mu\nu}(q,r,p)$ differ only at tree level; 
specifically, using  Eq.~(\ref{deco}), we see immediately that 
\be
\bqq_{\alpha\mu\nu}^{(0)}(q,r,p)  = \gtree_{\alpha\mu\nu}(q,r,p).  
\ee
Incidentally, notice that $\bqq_{\alpha\mu\nu}(q,r,p)$ coincides 
with the vertex appearing in diagram $(a_1)$ of the SDE~(\ref{gl-1ldr}), 
when projected to the Landau gauge~\cite{Aguilar:2008xm}, see also Section~\ref{cons}.  

\begin{figure}[!t]
\includegraphics[scale=.73]{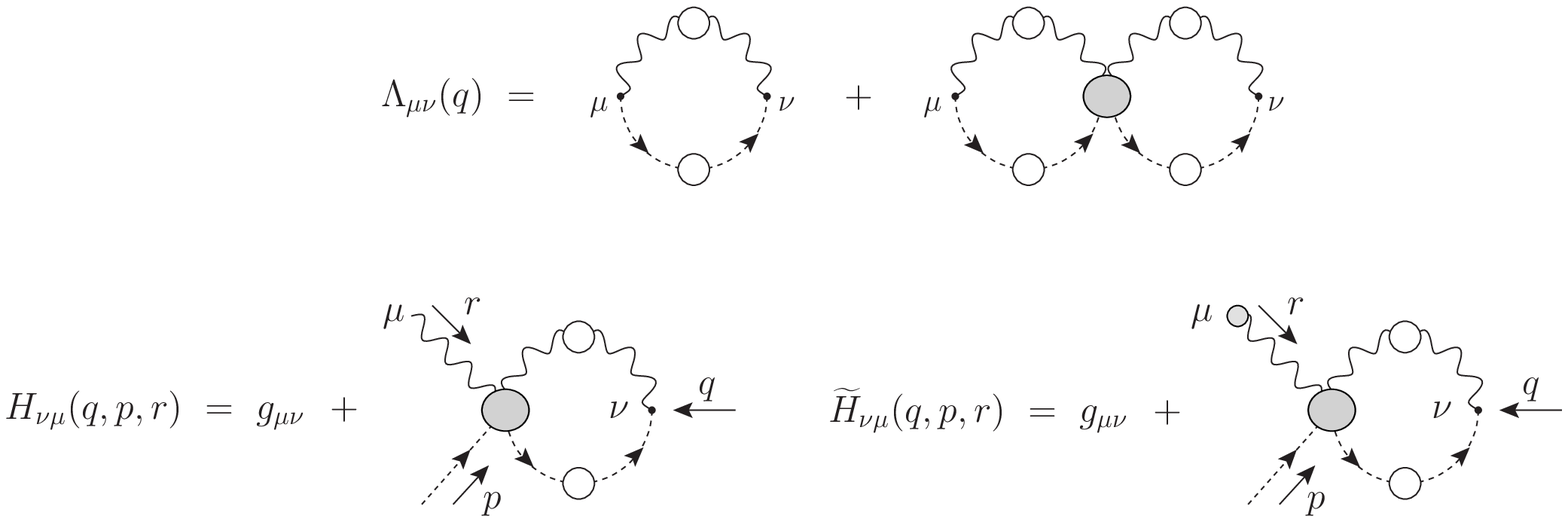}
\caption{\label{H-Htilde}Definitions and conventions of the auxiliary functions $\Lambda$, $H$ and $\widetilde{H}$. 
The color and coupling dependence for the combination shown, $c^a(q)A_\mu^b(r)A_\nu^{*c}(p)$, is $gf^{acb}$. 
White blobs represent connected Green's functions, while gray blobs denote  
one-particle irreducible (with respect to vertical cuts) kernels.}
\end{figure}

Then, the vertex $\bqq$ 
satisfies a (ghost-free) WI when contracted with the momentum $q_\alpha$ of the 
background gluon, whereas it satisfies a STI when contracted with 
the momentum of the quantum gluons ($r_\mu$ or $p_\nu$). They read
\bea
q^\alpha\bqq_{\alpha\mu\nu}(q,r,p)&=&p^2\bcj(p^2)P_{\mu\nu}(p)-r^2\bcj(r^2)P_{\mu\nu}(r)
\nonumber \\
r^\mu\bqq_{\alpha \mu \nu}(q,r,p)&=&F(r^2)\left[q^2\bcjb(q^2)P_\alpha^\mu(q)H_{\mu\nu}(q,r,p)-
p^2\bcj(p^2)P_\nu^\mu(p)\widetilde{H}_{\mu\alpha}(p,r,q)\right] \nonumber \\
p^\nu\bqq_{\alpha \mu \nu}(q,r,p)&=&F(p^2)\left[r^2\bcj(r^2)P_\mu^\nu(r)\widetilde{H}_{\nu\alpha}(r,p,q)-
q^2\bcjb(q^2) P_\alpha^\nu(q)H_{\nu\mu}(q,p,r)\right],
\label{STIs}
\eea
where $F(q^2)$ represents the ghost dressing function, related to the ghost propagator $D^{ab}(q^2)=\delta^{ab}D(q^2)$ through
\be
i D(q^2)= i \frac{F(q^2)}{q^2},
\ee
and the function $\bcjb$ is related to the conventional one defined in~(\ref{prop}) through the so-called 
``background quantum identity''~\cite{Grassi:1999tp, Binosi:2002ez}
\be
\bcjb(q^2)=\left[1+G(q^2)\right]\bcj(q^2).
\ee
Finally, the function $G$ appearing above is the metric form factor in the Lorentz decomposition of 
the auxiliary function $\Lambda$, defined as  
\bea
\Lambda_{\mu\nu}(q)&=&-ig^2C_A\int_k\!\Delta_\mu^\sigma(k)D(q-k)H_{\nu\sigma}(-q,q-k,k)\nonumber\\
&=&g_{\mu\nu}G(q^2)+\frac{q_\mu q_\nu}{q^2}L(q^2),
\eea
with $C_A$ the Casimir	eigenvalue of the adjoint	representation	[$C_A=N$ for $SU(N)$]. This function, together with the definitions and conventions for the auxiliary functions $H$ and $\widetilde{H}$, which will be studied in great detail in the next section, is shown in Fig.~\ref{H-Htilde}. Thus, requiring the vertex Ansatz to satisfy the STIs above implies that in its expression certain combinations of the ghost auxiliary functions $G$, $H$ and $\widetilde{H}$ will also appear. 

\section{\label{ghost}Identities of the ghost sector}

\noindent In view of the prominent role played by the ghost auxiliary functions $H$ and $\widetilde{H}$ in the ensuing analysis, 
in this section we shall study them, as well as the identities they satisfy. The framework that allows us to do 
this is the one developed long ago by Batalin and  Vilkovisky~\cite{Batalin:1977pb,Batalin:1981jr}, that we very briefly review below.

In the  Batalin-Vilkovisky formulation of Yang-Mills theories, 
one starts by introducing certain sources (called anti-fields and represented with a * super-index) and couples them to the corresponding field  through the term $\Phi^*s\,\Phi$ where $s$ is the BRST operator. Since these anti-fields will describe the renormalization of composite operators, they might be introduced only for those fields that transform non-linearly under the BRST operator; in the case of the $SU(N)$ Yang-Mills theories that we consider, this means for the gluon and ghost field only, since
\be
sA^a_\mu=({\cal D}^\mu c)^a; \qquad sc^a=-\frac12gf^{abc}c^bc^c
\ee
where ${\cal D}$ is the usual covariant derivative with $({\cal D}_\mu \Phi)^a=\partial_\mu \Phi^a+gf^{abc}A^b_\mu \Phi^c$. 

In much the same way, the quantization of the theory in a background field type of gauge requires, in addition to the aforementioned anti-fields, the introduction of new sources which couple to the BRST variation of the background fields~\cite{Grassi:1999tp}. These sources are sufficient for implementing the full set of symmetries at the quantum level, and, in the case of $SU(N)$ Yang-Mills theories, after choosing a linear gauge fixing function ({\it e.g.}, $R_\xi$ or BFM type of gauges), we are lead to the master equation
\be
\int\diff^4x\left[\frac{\delta\Gamma}{\delta A^{*\mu}_a}\frac{\delta\Gamma}{\delta A^{a}_\mu}+\frac{\delta\Gamma}{\delta c^{*a}}\frac{\delta\Gamma}{\delta c^a}+\Omega^\mu_a\left(\frac{\delta\Gamma}{\delta \widehat{A}^a_\mu}-\frac{\delta\Gamma}{\delta A^a_\mu}
\right)\right]=0.
\label{me}
\ee
In the formula above, $\Gamma$ is the (reduced) effective action, $A^*$ and $c^*$ the gluon and ghost anti-fields, $\widehat{A}$ 
is the gluon background field, and $\Omega$ the corresponding background source.

For  the study of the algebraic structure $H$ and $\widetilde{H}$, we will need two additional equations. 
The first one is the Faddeev-Popov equation, which controls the result of the contraction 
of an anti-field leg with the corresponding momentum. In position space, it reads
\be
\frac{\delta\Gamma}{\delta \bar c^a}+\left(\widehat{\cal D}^\mu\frac{\delta\Gamma}{\delta A^*_\mu}\right)^a-\left({\cal D}^\mu\Omega_\mu\right)^a-gf^{amn}\widehat{A}^m_\mu\Omega^\mu_n=0,
\label{FPE}
\ee
where $\widehat{\cal D}$ is the background covariant derivative (obtained from the usual one by replacing the 
gluon field $A$ with the background field $\widehat{A}$). Notice that Eqs.~(\ref{me}) and~(\ref{FPE}) 
above are given for the BFM gauge; to get the analogous expressions for the conventional $R_\xi$ gauges, 
one needs to set 
the background field and source $\widehat{A}$ and $\Omega$ to zero, and $\Gamma \to\left.\Gamma\right|_{\widehat{A},\Omega=0}$.

The second equation furnishes the WI functional ${\cal W}$, which encodes the residual background gauge invariance; it reads
\be
{\cal W}_{\vartheta}[\Gamma]=\int\!\diff^4x\!\sum_{\varphi=\Phi,\Phi^*}\!\left(\delta_{\vartheta}\varphi\right)\frac{\delta\Gamma}{\delta\varphi}=0,
\label{WI_gen_funct}
\ee 
where $\vartheta^a$ (which, in this case, plays the role of the ghost field) is the local infinitesimal parameter associated with  
the $SU(N)$ generators $t^a$; the local transformations of the fields are given by
\bea
\delta_{\vartheta}A^a_\mu=gf^{abc}A^b_\mu \vartheta^c &\qquad& \delta_{\vartheta}\widehat{A}^a_\mu=\partial_\mu \vartheta^a+gf^{abc}\widehat{A}^b_\mu \vartheta^c,\nonumber \\
\delta_{\vartheta} c^a=-g f^{abc}c^b\vartheta^c &\qquad& \delta_{\vartheta} \bar c^a=-g f^{abc}\bar c^b\vartheta^c . 
\label{theta_trans}
\eea
The anti-fields transformations coincide with those  
of the corresponding quantum fields given above, according to their specific representations.

After this detour, we are now in a position to study the ghost auxiliary functions in some depth.  
Let us start by introducing the notation
\bea
i\Gamma_{c^a A^b_\mu A^{*c}_\nu}(q,r,p)=igf^{acb}H_{\nu\mu}(p,q,r);&\qquad&
H^{(0)}_{\nu\mu}(p,q,r)=g_{\mu\nu}\nonumber \\
i\Gamma_{c^a \widehat{A}^b_\mu A^{*c}_\nu}(q,r,p)=igf^{acb}\widetilde{H}_{\nu\mu}(p,q,r);
&\qquad&
\widetilde{H}^{(0)}_{\nu\mu}(p,q,r)=g_{\mu\nu}.
\label{Hdef}
\eea
Then,  the ghost equation~(\ref{FPE}) allows to relate $H$ and $\widetilde{H}$ to the corresponding gluon-ghost vertices $\Gamma_{cA\bar c}$ and $\Gamma_{c\widehat{A}\bar c}$; indeed one has~\cite{Binosi:2008qk}
\bea
ip^\nu\Gamma_{c^b A^a_\mu A^{*c}_\nu}(r,q,p)+\Gamma_{c^b A^a_\mu \bar c^c}(r,q,p)&=&0\nonumber \\
ip^\nu\Gamma_{c^b \widehat{A}^a_\mu A^{*c}_\nu}(r,q,p)+\Gamma_{c^b \widehat{A}^a_\mu \bar c^c}(r,q,p)&=&igf^{cad}\Gamma_{c^b A^{*d}_\nu}(r).
\eea
Writing
\bea
i\Gamma_{c^b A^a_\mu \bar c^c}(r,q,p)=gf^{acb}\Gamma_{\mu}(r,q,p); &\qquad& \Gamma^{(0)}_{\mu}(r,q,p)=-p_\mu \nonumber \\
i\Gamma_{c^b \widehat{A}^a_\mu \bar c^c}(r,q,p)=gf^{acb}\widetilde{\Gamma}_{\mu}(r,q,p); &\qquad& \widetilde{\Gamma}^{(0)}_{\mu}(r,q,p)=(r-p)_\mu,
\eea
and using Eqs.~(\ref{fpe-rel-1}) and (\ref{Hdef}), we find
\bea
p^\nu H_{\nu\mu}(p,r,q)+\Gamma_{\mu}(r,q,p)&=&0 \nonumber \\
p^\nu \widetilde{H}_{\nu\mu}(p,r,q)+\widetilde{\Gamma}_{\mu}(r,q,p)&=&r_\mu F^{-1}(r^2).
\eea

\begin{figure}[!t]
\includegraphics[scale=.75]{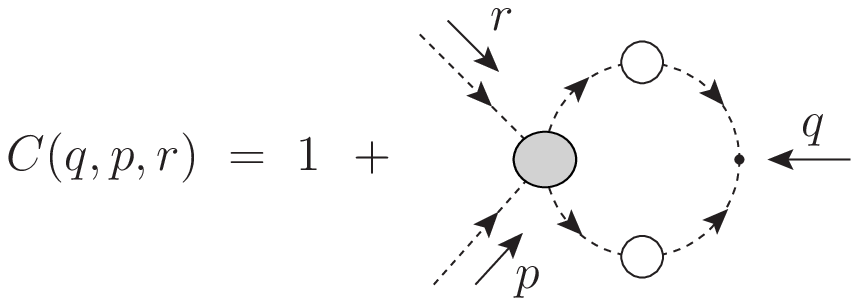}
\caption{\label{C}Definitions and conventions of the auxiliary function $C$. 
The color and coupling dependence for the combination shown, $c^a(q)c^b(r)c^{*c}(p)$, is $gf^{acb}$. The kernel is 
one-particle irreducible with respect to vertical cuts}
\end{figure}

As a second property, let us derive the WI satisfied by $\widetilde{H}$ when contracted with the momentum of the background gluon and the corresponding STI for $H$ when contracted by the momentum of the quantum gluon.
Starting from the functional derivative
\be
\left.\frac{\delta^3{\cal W}_\vartheta[\Gamma]}{\delta\theta^a(q)\delta c^b(r)\delta A^{*c}_\nu(p)}\right|_{\Phi,\Phi^*,\Omega=0};\qquad q+p+r=0,
\ee 
we get
\be
q^\mu\Gamma_{c^b\widehat{A}^a_\mu A^{*c}_\nu}(r,q,p)-gf^{bad}\Gamma_{c^d A^{*c}_\nu}(p)+gf^{dca}\Gamma_{c^b A^{*d}_\nu}(r)=0.
\label{HWI-1}
\ee
The ghost equation allows to relate the two-point function $\Gamma_{cA^*}$ to the ghost dressing function $F$ introduced before, through~\cite{Binosi:2008qk}
\be
\Gamma_{c^a A^{*b}_\nu}(q)=\delta^{ab}q_\nu F^{-1}(q^2),
\label{fpe-rel-1}
\ee
and, using this latter equation as well as the definition~(\ref{Hdef}), we can cast the identity~(\ref{HWI-1}) in its final form
\be
q^{\mu}\widetilde{H}_{\nu\mu}(p,r,q)=-p_\nu F^{-1}(p^2)-r_\nu F^{-1}(r^2).
\label{HWI}
\ee
Considering the functional derivative
\be
\left.\frac{\delta^3{\cal S}[\Gamma]}{\delta c^a(q)\delta c^b(r)\delta A^{*c}_\nu(p)}\right|_{\Phi,\Phi^*,\Omega=0};\qquad q+p+r=0,
\ee
one gets instead
\be
-\Gamma_{c^a A^{*\mu}_d}(q)\Gamma_{c^b A^d_\mu A^{*c}_\nu}(r,q,p)+
\Gamma_{c^bA^{*\mu}_d}(r)\Gamma_{c^a A^d_\mu A^{*c}_\nu}(q,r,p)-\Gamma_{c^ac^bc^{*c}}(q,r,p)\Gamma_{c^d A^{*c}_\nu}(p)=0.
\ee
Defining (see Fig.~\ref{C})
\be
i\Gamma_{c^ac^bc^{*c}}(q,r,p)=-igf^{acb}C(q,r,p),
\ee
and using the results~(\ref{Hdef}) and (\ref{fpe-rel-1}), we finally get
\be
q^{\mu}H_{\nu\mu}(p,r,q)=-F(q^2)\left[p_\nu F^{-1}(p^2) C(q,r,p)+r^\mu F^{-1}(r^2)H_{\nu\mu}(p,q,r)\right].
\label{HSTI}
\ee

To proceed further, we decompose the auxiliary functions $H$ and $\widetilde{H}$ in terms of their basic tensor forms  
\bea
H_{\nu\mu}(p,r,q)&=&g_{\mu\nu}a_{qrp}-r_\mu q_\nu b_{qrp}+q_\mu p_\nu c_{qrp}+q_\nu p_\mu d_{qrp}+p_\mu
p_\nu e_{qrp} \nonumber \\
\widetilde{H}_{\nu\mu}(p,r,q)&=&g_{\mu\nu}\widetilde{a}_{qrp}-r_\mu q_\nu \widetilde{b}_{qrp}+q_\mu p_\nu \widetilde{c}_{qrp}+q_\nu p_\mu \widetilde{d}_{qrp}+p_\mu p_\nu \widetilde{e}_{qrp},
\label{Hdec}
\eea
where, following the notation of~\cite{Ball:1980ax} we have introduced the shorthand notation 
$a_{qrp}$ for $a(q,r,p)$, and similarly for all other form factors appearing in (\ref{Hdec}). Then, one can use the identities (\ref{HWI}) and (\ref{HSTI}) in order to constrain certain combinations of these form factors. Indeed from the WI~(\ref{HWI}) one finds
\bea
\widetilde{a}_{qrp}-(q\cdot r)\widetilde{b}_{qrp}+(q\cdot p)\widetilde{d}_{qrp}&=&F^{-1}(r^2)\nonumber \\
q^2\widetilde{c}_{qrp}+(q\cdot p)\widetilde{e}_{qrp}+F^{-1}(p^2)&=&F^{-1}(r^2),
\label{Hhatconstr}
\eea
while the STI gives
\bea
F(r^2)\left[a_{qrp}-(q\cdot r)b_{qrp}+(q\cdot p)d_{qrp}\right]&=&F(q^2)\left[a_{rqp}-(q\cdot r)b_{rqp}+(p\cdot r)d_{rqp}\right] \nonumber \\
F^{-1}(q^2)\left[q^2c_{qrp}+(q\cdot p) e_{qrp}\right]+F^{-1}(p^2)C_{qrp}&=&F^{-1}(r^2)\left[a_{rqp}-(q\cdot r)b_{rqp}  +(p\cdot r)d_{rqp} \right]\nonumber \\
&-&F^{-1}(r^2)\left[r^2 c_{rqp}+ (p\cdot r)e_{rqp}\right],
\label{Hconstr}
\eea
where, as before, $C_{qrp} \equiv C(q,r,p)$.

The first equation of (\ref{Hconstr}), together with those obtained through cyclic permutations of momenta 
and indices, represent the aforementioned constraints, first  
found in~\cite{Ball:1980ax} [{\it viz.} Eq.~(2.10) in that article] as necessary conditions for solving the STIs of the $QQQ$ vertex. 
It is clear from the above analysis that these constraints are a direct consequence of the STI satisfied 
by the function $H$ (in~\cite{Ball:1980ax} their validity was explicitly verified at the one-loop level only). 

Finally, let us conclude this section by observing that $H$ and $\widetilde{H}$ can be related.  Specificcally, the functional differentiation
\be
\left.\frac{\delta^3{\cal S}[\Gamma]}{\delta\Omega^a_\alpha(q)\delta c^b(r)\delta A^{*c}_\nu(p)}\right|_{\Phi,\Phi^*,\Omega=0};\qquad q+p+r=0,
\ee
furnishes the corresponding BQI, namely 
\bea
{i}\Gamma_{c^b\widehat{A}^a_\mu A^{*c}_\nu}(r,q,p)&=&\left[{i}g_\mu^\rho\delta^{ad}+\Gamma_{\Omega^a_\mu A^{*\rho}_d}(q)\right]\Gamma_{c^b A^d_\rho A^{*c}_\nu}(r,q,p)+\Gamma_{c^b A^{*\rho}_d}(r)
\Gamma_{\Omega^a_\alpha A_\rho^d A^{*c}_\nu}(q,r,p)\nonumber \\
&-&\Gamma_{\Omega^a_\mu c^b c^{*d}}(q,r,p)\Gamma_{c^d A^{*c}_\nu}(p).
\eea

\section{Solving the Ward and Slavnov-Taylor identities}
\label{syst}

In this section we proceed to the actual construction of the vertex $\bqq$,  
by solving the WI and STIs given in Eq.~(\ref{STIs}). 

In order to simplify the resulting equations, it is convenient to follow~\cite{Ball:1980ax} 
and group the 14 possible tensor forms into two sets corresponding to the 
longitudinal and the (totally) transverse parts of the vertex. 
One begins by decomposing the vertex according to
\be
\bqq^{\alpha\mu\nu}(q,r,p)=\bqq_{(\ell)}^{\alpha\mu\nu}(q,r,p)+\bqq_{(t)}^{\alpha\mu\nu}(q,r,p).
\label{decomp}
\ee
The longitudinal part is then characterized by 10 form factors $X_i$ according to
\be
\bqq_{(\ell)}^{\alpha\mu\nu}(q,r,p)=\sum_{i=1}^{10}X_i(q,r,p)\ell_i^{\alpha\mu\nu}(q,r,p),
\label{tenlon}
\ee
with the explicit form of the tensors $\ell^i$ given by
\be
\begin{tabular}{lll}
$\ell_1^{\alpha\mu\nu} =  (q-r)^{\nu} g^{\alpha\mu}$
& 
$\ell_2^{\alpha\mu\nu} =  - p^{\nu} g^{\alpha\mu}$\hspace{.75cm}
&
$\ell_3^{\alpha\mu\nu} =  (q-r)^{\nu}[q^{\mu} r^{\alpha} -  (q\cdot r) g^{\alpha\mu}] $\\
$\ell_4^{\alpha\mu\nu} = (r-p)^{\alpha} g^{\mu\nu}$
&
$\ell_5^{\alpha\mu\nu} =  - q^{\alpha} g^{\mu\nu}$
&
$\ell_6^{\alpha\mu\nu} =  (r-p)^{\alpha}[r^{\nu} p^{\mu} -  (r\cdot p) g^{\mu\nu}]$
\\
$\ell_7^{\alpha\mu\nu} =  (p-q)^{\mu} g^{\alpha\nu}$
&
$\ell_8^{\alpha\mu\nu} = - r^{\mu} g^{\alpha\nu}$
&
$\ell_9^{\alpha\mu\nu} = (p-q)^{\mu}[p^{\alpha} q^{\nu} -  (p\cdot q) g^{\alpha\nu}]$
\\
$\ell_{10}^{\alpha\mu\nu} = q^{\nu}r^{\alpha}p^{\mu} + q^{\mu}r^{\nu}p^{\alpha}$. & &
\end{tabular}
\label{Ls}
\ee
Notice that excluding $\ell_{10}$, each of the remaining $\ell_{i+3}$ can be obtained by the corresponding $\ell_i$ through cyclic permutation of momenta and indices; in addition,  Bose symmetry with respect to the quantum legs requires that $\bqq$ reverses sign under the interchange of  the corresponding Lorentz indices and  momenta, thus implying the relations
\be
\begin{tabular}{lll}
$X_1 (q,p,r) =  X_7 (q,r,p)$\hspace{0.75cm}
&
$X_2 (q,p,r) =   - X_8 (q,r,p)$\hspace{0.75cm}
&
$X_3 (q,p,r) =   X_9 (q,r,p)$
\\
$X_4 (q,p,r) =  X_4 (q,r,p)$
&
$X_5 (q,p,r) =  - X_5 (q,r,p)$
&
$X_6 (q,p,r) =   X_6 (q,r,p)$
\\
$X_{10} (q,p,r) =   - X_{10} (q,r,p),$ & &
\end{tabular}
\label{boserel}
\ee
which reduce the number of possible independent form factors from the original 10 to only~7.

The (undetermined) transverse part of the vertex is finally described by the remaining 4 form factors $Y_i$
\be
\bqq_{(t)}^{\alpha\mu\nu}(q,r,p)=\sum_{i=1}^{4}Y_i(q,r,p)t_i^{\alpha\mu\nu}(q,r,p),
\label{vtr}
\ee
with the completely transverse tensors $t^i$ given by
\bea
t_1^{\alpha\mu\nu} &=&  
[(q\cdot r) g^{\alpha\mu} - q^{\mu}  r^{\alpha}]
[(r\cdot p) q^{\nu} - (q\cdot p) r^{\nu}]
\nonumber\\
t_2^{\alpha\mu\nu} &=&  
[(r\cdot p) g^{\mu\nu} - r^{\nu}p^{\mu}]
[(p\cdot q) r^{\alpha} - (r\cdot q) p^{\alpha}]
\nonumber \\
t_3^{\alpha\mu\nu} &=&  
[(p\cdot q) g^{\nu\alpha} - p^{\alpha}q^{\nu}]
[(q\cdot r) p^{\mu} - (r\cdot p) q^{\mu}]
\nonumber\\
t_4^{\alpha\mu\nu}&=&g^{\mu\nu}[ (p\cdot q)r^\alpha-(r\cdot q)p^\alpha ]+g^{\alpha\mu}[(r\cdot p)q^\nu-(q\cdot p)r^\nu ] +g^{\alpha\nu}[(r\cdot q)p^\mu -(r\cdot p)q^\mu]\nonumber \\
&+&p^\alpha q^\mu r^\nu-r^\alpha p^\mu q^\nu.
\label{Ts}
\eea

The form factors $X_i$ are then fully determined 
by solving the system of linear equations generated by the identities given in Eq.~(\ref{STIs}). 
The procedure is conceptually straightforward, but operationally rather cumbersome. 
One first substitutes on the lhs of Eq.~(\ref{STIs}) the general tensorial decomposition of  
$\bqq^{(\ell)}$ given in Eq.~(\ref{tenlon}), and then equates 
the coefficients of the resulting tensorial structures to those appearing on the rhs. 
Thus, one obtains a system of equations expressing the form factors $X_i$ 
in terms of combinations of quantities such as $J$, $F$, {\it etc}. 

In what follows we will only report the set of independent equations, {\it i.e.}, we will omit
equations that can be obtained from existing ones by implementing the change $p \leftrightarrow r$ 
and using the constraints of ~(\ref{boserel}). Thus, for example, the equation 
$X_7 + X_8 + (q\cdot r)X_{10} =  J(p)$ does not form part of the set of independent equations, 
because it can be obtained from the second equation in Eq.(\ref{absys}) below, by carrying out the 
aforementioned transformation, and using the corresponding relations from Eq.~(\ref{boserel}).

Thus, from the Abelian WI one obtains the following 4 equations  
\bea
(p^2-r^2) X_4  - q^2 X_5 - (r\cdot p) (p^2-r^2)X_6 &=&  p^2J(p) - r^2J(r)
\nonumber\\
X_1 - X_2 - (q\cdot p)X_{10} &=&   J(r)
\nonumber\\
X_1 + X_2 - X_7 + X_8 &=&  0
\nonumber\\
2 X_1 + (p^2-r^2) X_6 - 2 X_7  + q^2 X_{10} &=&  0,
\label{absys}
\eea
where the form of the second equation has been simplified by making use of the third.

Similarly, from the non-Abelian STI one obtains 
\bea
(r^2-q^2) X_1  - p^2 X_2 - (q\cdot r) (r^2-q^2)X_3 &=&  F(p)
\left[{\widetilde a}_{qpr} r^2 J(r) - a_{rpq} q^2{\widetilde J}(q)\right]
\nonumber\\
(r^2-q^2) X_3 - 2  X_4 + 2 X_7 + p^2 X_{10} &=& F(p)
\left[({\widetilde b}_{qpr}+ {\widetilde d}_{qpr}) r^2 J(r) 
- (b_{rpq}+d_{rpq}) q^2{\widetilde J}(q)\right]
\nonumber\\
- X_7 +  X_8 + (r\cdot p) X_{10} &=&   F(p)
\left\{(r\cdot p) {\widetilde b}_{qpr} J(r) 
- \left[a_{rpq} +(q\cdot r){d}_{rpq}\right]{\widetilde J}(q)
\right\}
\nonumber\\
X_4 + X_5 + (q\cdot p)X_{10} &=& F(p)
\left\{\left[{\widetilde a}_{qpr} +(q\cdot r) {\widetilde d}_{qpr}\right] J(r) 
- (q\cdot p) b_{rpq} {\widetilde J}(q)\right\}
\nonumber\\
-X_4 + X_5 + X_7 + X_8  &=&  (q\cdot r)F(p)
\left[-{\widetilde b}_{qpr} J(r) 
+ b_{rpq} {\widetilde J}(q)\right].
\label{nabsys}
\eea
Clearly, there are 5 additional equations, obtained from the second STI; however, they too  
can be obtained from the set of equations~(\ref{nabsys})
by imposing the transformation $r \leftrightarrow p$ and using the 
relations~(\ref{boserel}), and are therefore omitted. 

Eqs.~(\ref{absys}) and~(\ref{nabsys}) furnish a set of 9 equations for the 7 independent longitudinal form factors of~Eq.~(\ref{boserel}); therefore the existence of a (unique) solution to this system, requires the appearance of 2 non-trivial constraints for the ghost sector which read
\bea
& & F(r^2) [{a}_{prq} - (r\cdot p){b}_{prq} + (q\cdot p){d}_{prq}]
= F(p^2)[{a}_{rpq} - (r\cdot p){b}_{rpq} + (q\cdot r){d}_{rpq}]\nonumber \\
& &F(r^2) [{\widetilde a}_{qrp} - (q\cdot r){\widetilde b}_{qrp} + (q\cdot p){\widetilde d}_{qrp}] = 1. 
\eea
Evidently these relations are 
nothing but an expression of the STI and the WI that the ghost auxiliary functions $H$ and $\widetilde{H}$ 
are bound to satisfy, as shown in Eqs.~(\ref{Hhatconstr}) and~(\ref{Hconstr}). 
Therefore the system can be solved and one finds a solution of the type presented in~\cite{Ball:1980ax} 
with a modified ghost-sector, reading
\bea
X_1(q,r,p) &=&  \frac{1}{4}{\widetilde J}(q) \left\{ 
- p^2 b_{prq} F(r) + 
[2 a_{rpq} + p^2 b_{rpq} + 2 (q\cdot r) d_{rpq} ]F(p) \right\}    
\nonumber\\
&+& \frac{1}{4} J(r)\left[ 2 +  (r^2-q^2) {\widetilde b}_{qpr} F(p)\right]
+ \frac{1}{4} J(p)\, p^2 \,{\widetilde b}_{qrp} F(r) 
\nonumber\\
X_2(q,r,p)  &=&  \frac{1}{4}{\widetilde J}(q)\left\{
(q^2- r^2) b_{prq} F(r) + 
[2 a_{rpq} + (r^2-q^2)b_{rpq} + 2 (q\cdot r) d_{rpq} ] F(p)\right\}  
\nonumber\\
&+& \frac{1}{4} J(r)\left[ - 2 +  p^2 {\widetilde b}_{qpr} F(p)\right]
+ \frac{1}{4} J(p)\, (r^2- q^2)\,{\widetilde b}_{qrp} F(r) 
\nonumber\\
X_3(q,r,p) &=& \frac{F(p)}{q^2-r^2} 
\left\{{\widetilde J}(q)\left[a_{rpq} - (q\cdot p) d_{rpq} \right] 
-  J(r) \left[{\widetilde a}_{qpr} - (r\cdot p){\widetilde d}_{qpr}\right] \right\}
\nonumber\\
X_4(q,r,p) &=& \frac{1}{4}{\widetilde J}(q) q^2 \left[b_{prq} F(r) + b_{rpq} F(p)\right]
+ \frac{1}{4} J(r)  \left[2 - q^2 {\widetilde b}_{qpr} F(p)\right]
\nonumber\\
&+& \frac{1}{4} J(p) \left[2 - q^2 {\widetilde b}_{qrp} F(r)\right]
\nonumber\\
X_5(q,r,p) &=& \frac{1}{4}{\widetilde J}(q) (p^2-r^2) \left[b_{prq} F(r) + b_{rpq} F(p)\right]
+ \frac{1}{4} J(r)  \left[2 +(r^2-p^2) {\widetilde b}_{qpr} F(p)\right]
\nonumber\\
&-& \frac{1}{4} J(p) \left[2 +(p^2-r^2) {\widetilde b}_{qrp} F(r)\right]
\nonumber\\
X_6(q,r,p) &=& \frac{J(r)-J(p)}{r^2-p^2}
\nonumber\\
X_7(q,r,p) &=& X_1(q,p,r)
\nonumber\\
X_8(q,r,p) &=& - X_2(q,p,r)
\nonumber\\
X_9(q,r,p) &=& X_3(q,p,r)
\nonumber\\
X_{10}(q,r,p) &=& \frac{1}{2}\left\{ 
{\widetilde J}(q) \left[b_{prq} F(r) - b_{rpq} F(p)\right]
+   J(r) F(p) {\widetilde b}_{qpr} - J(p) F(r) {\widetilde b}_{qrp} \right\}.
\label{X10}
\eea

Notice finally that from the above result one can obtain also the solution 
for the fully Bose-symmetric PT vertex $\widehat{\bm{\Gamma}}$, 
namely the $BBB$ vertex  originally constructed in~\cite{Cornwall:1989gv}, and further studied in~\cite{Binger:2006sj}. 
This vertex satisfies (with respect to any one of its three-legs) 
the WI shown in the first line of~(\ref{STIs}), 
with the modification $\Delta^{-1} \to \widehat{\Delta}^{-1}$, where 
\be
\widehat{\Delta}^{-1}(q^2)=q^2\widehat{J}(q^2);\qquad \widehat{J}(q^2)=[1+G(q^2)]^2J(q^2), 
\ee
is the inverse of the full PT-BFM gluon propagator.  
Thus, from the STIs appearing in Eq.~(\ref{STIs}) we see that the expression for the (longitudinal part) of the $BBB$ vertex 
(given in~\cite{Binger:2006sj}) may be recovered from Eq.~(\ref{X10})
by setting $J,\widetilde{J}\to\widehat{J}$, 
$F=1$ and $a=\widetilde{a}=1$, and all remaining form factors of $H$ and  $\widetilde{H}$ equal to zero.

\section{Consequences for the SDE of the gluon propagator}
\label{cons}

\noindent As has already been mentioned in previous sections, the Ansatz for the longitudinal part of the 
$BQQ$ vertex, obtained by ``solving'' the WI and STI that this vertex satisfies,  
is of central importance for the self-consistent treatment of the SDE equation 
governing the dynamics of the gluon self-energy. 
This fact may be best appreciated in the context of the SDE governing the 
gluon propagator~(\ref{gl-1ldr}) projected onto the Landau gauge, which is known to 
display a variety of field-theoretic subtleties.

In particular, a crucial self-consistency condition for the mechanism of dynamical 
gluon mass generation developed in a series of articles~\cite{Cornwall:1981zr,Aguilar:2006gr,Binosi:2007pi,
Binosi:2008qk,Aguilar:2008xm}
is the cancellation of all seagull-type of divergences, {\it i.e.}, 
divergences produced by integrals of the type $\int_k \Delta(k)$, or variations thereof~\cite{Aguilar:2009ke}.
In the case of the 
dimensional regularization that we use throughout, the presence of such integrals would give rise to 
divergences of the type $m_0^{2} (1/\epsilon)$, where  $m_0$ is the value of the dynamically generated gluon mass at $q^2=0$, 
{\it i.e.}, $m_0 = m(0)$; 
if a hard cutoff $\Lambda$ were to be employed, 
these latter terms would diverge quadratically, as $\Lambda^2$.
The disposal of such divergences would require the introduction in the original 
Lagrangian of a 
counter-term of the form $m^2_0 A^2_{\mu}$, which is, however, forbidden by the local gauge invariance, which  
must remain intact.

This is a point of paramount importance. Indeed, in the picture put forth
in the aforementioned articles, 
the Lagrangian of the Yang-Mills theory (or that of QCD) is never altered;  
the generation of the gluon mass takes place dynamically, 
without violating any of the underlying symmetries. Amplifying this point further, let us mention that,  
given that the  Lagrangian is never altered, the only other possible way of violating the gauge (or BRST) symmetry 
would be by not respecting, at some intermediate step, some of the WIs and STIs satisfied by the 
Green's functions involved; for example, in the conventional SDE formulation, a naive truncation would 
compromise the transversality of the resulting gluon self-energy, {\it i.e.}, the text-book condition 
$q^{\mu}\Pi_{\mu\nu}(q)=0$ would be no longer valid.  

Returning to the aforementioned seagull-type of divergences, 
as has been shown in detail in~\cite{Aguilar:2009ke}, their cancellation proceeds by means of the 
identity
\be
\int_k\! k^2\frac{\partial{\Delta}(k^2)}{\partial k^2}+\frac{\dim}2\int_k\!\Delta(k^2)=0,
\label{seagull}
\ee 
whose validity hinges on the special rules of dimensional regularization. 
In fact, as explained 
in ~\cite{Aguilar:2009ke}, in scalar QED it is exactly this 
identity that enforces the masslessness of the photon 
both perturbatively (at the level of a one-loop calculation) as well as 
non-perturbatively, at the level of the one-loop dressed SDE 
(assuming that the Schwinger mechanism is not in operation).  
In this context, the difference between scalar QED and Yang-Mills 
is that in the former case  $\Delta$ should be replaced by the propagator of the charged scalar field ${\cal D}$
entering into the loop of the photon self-energy, whereas in the latter, $\Delta$ is the gluon propagator itself 
(entering in its own gluonic loop, see Fig.~\ref{gSDE}).  
Note that the two types of 
integrals appearing on the lhs  of Eq.~(\ref{seagull}) are individually non-vanishing (in fact, they both diverge); 
it is only when they come in the particular 
combination shown above that they sum up to zero.

The difficulty associated with 
Eq.~(\ref{seagull}) is not so much recognizing its validity, but rather, having it triggered 
at the end of the calculation. 
Specifically, the ingredients entering into the SDE (most importantly, the vertex) 
must be such that, 
after taking the limit of the 
SDE as $q\to 0$, all seagull-type contributions must conspire to 
appear in the combination given on the lhs of Eq.~(\ref{seagull}) only!
In fact, the slightest change in a relative numerical factor will invalidate the entire construction.

Let us now see in detail how this seagull cancellation proceeds in the case of the 
Landau gauge SDE, supplied with the $BQQ$ vertex constructed earlier. 
The gluon self-energy obtained from the PT-BFM one-loop dressed SDE (see Fig.~\ref{gSDE}) in the Landau gauge 
reads   
\be
{\widehat\Pi}^{\mu\nu}(q) = g^2 C_A \sum_{i=1}^{5} A^{\mu\nu}_{i}(q),
\ee
with    
\bea
A^{\mu\nu}_{1}(q)&=& \frac{1}{2}\int_k
\Gamma^{(0)\mu}_{\alpha\beta}\Delta^{\alpha\rho}_{t}(k)\Delta^{\beta\sigma}_{t}(k+q)\bqq^\nu_{\rho\sigma}
\nonumber \\
A^{\mu\nu}_{2}(q)&=&  \int_k \!\Delta^{\alpha\mu}_{t}(k) \frac{(k+q)^{\beta} \Gamma^{(0)\nu}_{\alpha\beta}}{(k+q)^2}
\nonumber \\
A^{\mu\nu}_{3}(q)&=& \int_k \!\Delta^{\alpha\mu}_{t}(k) \frac{(k+q)^{\beta} \bqq^\nu_{\alpha\beta}}{(k+q)^2}
\nonumber \\
A^{\mu\nu}_{4}(q)&=&-\frac{(\dim-1)^2}{\dim}g^{\mu\nu}\int_k \Delta(k) \nonumber \\
A^{\mu\nu}_{5}(q)&=& \int_k \frac{k^{\mu}(k+q)^{\nu}}{k^2 (k+q)^2},
\label{thebees}
\eea
and $\Delta_{t}^{\mu\nu}(q)=P^{\mu\nu}(q)\Delta(q^2)$ is the transverse Landau gauge propagator.

First of all, it is rather straightforward to verify explicitly that, if $\bqq$ satisfies the WI of Eq.~(\ref{STIs}),   
\be
q^{\mu} {\widehat\Pi}_{\mu\nu}(q) =0 .
\ee
Therefore, ${\widehat\Pi}^{\mu\nu}(q) = P^{\mu\nu}(q)\widehat{\Pi}(q^2)$, and the scalar function  $\widehat{\Pi}(q^2)$ is given by 
\be
\widehat{\Pi}(q^2) = \frac{g^2 C_A}{\dim-1} \sum_{i=1}^{5} A_{i\,\mu}^{\mu}(q) . 
\label{pisc}
\ee

Since we are interested in the behavior of $\widehat{\Pi}(0)$, and in particular the annihilation of any seagull-type of divergence, 
we next take the limit  $q\to0$ of the rhs of Eq.~(\ref{pisc}), 
using the explicit form of the vertex $\bqq$ derived in the preceding section.
In addition, we will assume that all form factors appearing in the Lorentz decomposition of $H$ and $\widetilde{H}$ 
are regular in the $q\to0$ limit; this is a reasonable assumption, 
given that the generation of a dynamical mass is expected to regulate all potential infrared divergences.

Consider then the term $A_1$; after replacing $r\to k $ and $p\to-k-q$, 
and dropping terms proportional to $k_ \rho$ and $(k+q)_\sigma$, given that they vanish when contracted with the 
corresponding transverse propagators, we find that the tensor structures $\ell_i$ are such that
\be
\ell_1^{\beta\rho\sigma}\sim\ell_5^{\beta\rho\sigma}\sim\ell_7^{\beta\rho\sigma}\sim{\cal O}(q); \qquad
\ell_3^{\beta\rho\sigma}\sim\ell_9^{\beta\rho\sigma}\sim{\cal O}(q^2); \qquad
\ell_{10}^{\beta\rho\sigma}\sim{\cal O}(q^3).
\ee
Since $\ell_2$ and $\ell_8$ yield a vanishing result 
when contracted with the transverse propagators, the only tensors surviving will be $\ell_4$ and $\ell_6$; after taking the trace, 
one is then left with the result
\be
A_1=2(\dim-1)\int_k\!k^2\Delta(k)\Delta(k+q)\left[X_4 +k\cdot(k+q)X_6 \right]+{\cal O}(q).
\ee
On the other hand, using the explicit results given in  Eq.~(\ref{X10}), one has
\be
X_4+k\cdot(k+q)X_6=\frac{\Delta^{-1}(k+q)-\Delta^{-1}(k)}{(k+q)^2-k^2}-\frac{q^2}{2} \frac{J(k+q)-J(k)}{(k+q)^2-k^2},
\ee
and  the second term is easily shown to vanish as $q$ goes to zero. Thus, one is finally left with the result
\be
A_1=-2(\dim-1)\int_k\!k^2\frac{\Delta(k+q)-\Delta(k)}{(k+q)^2-k^2}+{\cal O}(q).
\ee
Thus, in the limit $q\to0$ one obtains
\be
\frac{\Delta(k+q)-\Delta(k)}{(k+q)^2-k^2}= \frac{\partial{\Delta}(k^2)}{\partial k^2}+{\cal O}(q^2),
\ee 
which gives rise to  the first term on the lhs of Eq.~(\ref{seagull}).

The second contribution to  Eq.~(\ref{seagull}) arises from two terms, the obvious term $A_4$, which already has the 
required form (but not the right numerical coefficient), and the term $A_2$, 
which, after letting the momentum $(k+q)^{\beta}$ act on the tree-level vertex $\Gamma^{(0)\nu}_{\alpha\beta}$, 
reads 
\be
A_2=-(\dim-1)\int_k\Delta(k)+{\cal O}(q).
\ee
Putting all pieces together, we finally obtain
\bea
A_1+A_2+A_4&=&-2\left[\int_k\!k^2\frac{\Delta(k+q)-\Delta(k)}{(k+q)^2-k^2}+\frac{\dim}2\int_k\!\Delta(k)\right]+\mathcal{O}(q)\nonumber \\
&\stackrel{q\to0}{\longrightarrow}&-2\left[\int_k\! k^2\frac{\partial{\Delta}(k^2)}{\partial k^2}+\frac{\dim}2\int_k\!\Delta(k)\right]=0,
\label{canc}
\eea
that is, one recovers the seagull cancellation condition of Eq.~(\ref{seagull}).

Let us finally look at what happens to the remaining terms $A_3$ and $A_5$ as $q\to0$. 
To begin with, it is elementary to check that $A_5(0) =0$.
The treatment of $A_3$ is more subtle, and makes manifest the need 
to satisfy the STI of Eq.~(\ref{STIs}).  
Specifically, after taking the trace and using Eq.~(\ref{STIs}), one finds 
\bea
A_3&=&-\int_k\!\Delta_{t}^{\mu\rho}(k)\frac{F(k+q)}{(k+q)^2}\bigg[\Delta^{-1}(k)P_\rho^\sigma(k)\widetilde{H}_{\sigma\mu}(k,-k-q,q)
\nonumber \\
&-&\widetilde{\Delta}^{-1}(q)P^\sigma_\rho(q)H_{\sigma\mu}(q,-k-q,k)\bigg].
\label{someq}
\eea
Substituting the expansion for $\widetilde{H}$ given in Eq.~(\ref{Hdec}), the first term on the rhs of Eq.~(\ref{someq})
yields
\be
-\int_k\!\frac{F(k+q)}{(k+q)^2}P^{\mu\sigma}(k)\left[g_{\mu\sigma}\widetilde{a}+(k+q)_\mu q_\sigma\widetilde{b}\right],
\ee
where the argument of the form factors $\widetilde{a}$ and $\widetilde{b}$ is now $(q,-k-q,k)$. 
In the $q\to0$ limit this term vanishes in dimensional regularization, by virtue of the well-known property  
$\int_k k^{-2} =0$. Indeed, 
\be
\int_k\!\frac{F(k)}{k^2}\widetilde{a}(0,-k,k)=\int_k\!\frac1{k^2}=0,
\ee
where we have used the fact that, in this limit, the first equation of~(\ref{Hhatconstr}) reduces to the simple relation  
\be
\widetilde{a}(0,-k,k)=F^{-1}(k).
\ee

Next, consider the second term in Eq.~(\ref{someq}); using now the expansion~(\ref{Hdec}) for $H$, we obtain 
\be
\widetilde{\Delta}^{-1}(q)\int_k\!\Delta(k)\frac{F(k+q)}{(k+q)^2}P_\mu^\rho(k)P^{\mu\sigma}(q)\left[g_{\rho\sigma}a-(k+q)_\rho k_\sigma b+k_\sigma q_\rho d\right],
\ee
where now the form factors $a$, $b$ and $d$ carry the argument $(k,-k-q,q)$. 
In the $q\to0$ limit this integral gives a finite term, proportional to the expression
\be
\int_k\!\Delta(k)\frac{F(k)}{k^2}a(k,-k,0).
\label{A32}
\ee
It should be noticed that 
the contribution of the term $A_3$ is rendered finite
precisely by virtue of the special properties of the ghost sector. 
In fact, if one were to use in the SDE 
the fully Bose-symmetric PT vertex $BBB$ (instead of 
the correct $BQQ$ vertex that we use here),  
the term $A_3$ would give rise to an ultraviolet divergence, since 
(as explained at the end of the previous section) one should set 
in (\ref{A32}) $F,a\to1$.

These observations demonstrate that, as happens in the case of chiral symmetry breaking ~\cite{Aguilar:2010cn},   
the complete treatment of the ghost dynamics is instrumental also 
for the self-consistency of the mass generation in the purely gluonic sector of QCD.

\section{\label{concl}Conclusions}

\noindent  In  this article  we  have  constructed  a gauge  technique
inspired Ansatz for the $BQQ$ three-gluon vertex $\bqq$ that naturally
arises in  the context of the  PT-BFM derivation of  the SDE equations
for  Yang-Mills theories.   An indispensable  step for  realizing this
construction   has    been   the   formal    derivation   within   the
Batalin-Vilkovisky  formalism of the  all-order WI  (respectively STI)
satisfied by the ghost Green's functions $\widetilde{H}$ (respectively
$H$),   which,  as  shown   in  Section~\ref{syst},   furnish  crucial
constraints that allow  $\bqq$ to conform with both the  WI as well as
the STIs of Eq.~(\ref{STIs}).

It  is important  to emphasize  that  the analysis  presented here  is
completely  general, and  in  particular that  the  solution shown  in
Eq.~(\ref{X10})  is   valid  irrespectively   of  the  value   of  the
gauge-fixing parameter  used to quantize  the theory. To be  sure, the
various ingredients  appearing in  Eq.~(\ref{X10}), such as  $J$, $F$,
{\it etc.}, depend explicitly  on $\xi$ (or on $\xi_Q$); nevertheless,
the precise functional dependence of the form factors $X_{i}$ on these
functions, is always valid, given that it originates from the solution
of the  WI and STIs~(\ref{STIs}), whose  form is in  turn gauge fixing
parameter  independent.   This  is  particularly  relevant,  given  the
existing   perspectives~\cite{Cucchieri:2009kk,Cucchieri:2011aa}    of   
carrying   out
large-volume lattice simulations of the gluon and ghost propagators in
covariant gauges other than the  Landau gauge, {\it i.e.}, at $\xi\neq
0$.  In  particular,  the  possibility of  simulating  propagators  in
background-type  gauges  (especially  the  background  Feynman  gauge,
$\xi_Q=1$) opens up the interesting prospect of studying central 
quantities of the PT-BFM approach directly on  the lattice~\cite{Cucchieri:2011aa}.

As  already mentioned in the Introduction, the construction based on 
solving the WI and STI leaves the 
transverse part of the vertex undetermined. In terms of the notation 
introduced in section~\ref{syst}, this means that the four form factors 
$Y_i$ appearing in  Eq.~(\ref{vtr}) are completely unconstrained. 
In the case of QED, it is known that, in the presence of a mass gap, 
the transverse part of the photon-electron vertex is sub-leading in the 
infrared. Even though we are not aware of a similar study 
in a non-Abelian context, it is reasonable to assume that 
this will continue to be so, provided that a mass gap 
({\it i.e.}, dynamical gluon mass) has indeed been generated. In such a case,  
one would expect that the  identically  conserved  part of the vertex  
should vanish more rapidly by at least one power of
$q$ compared to the  longitudinal part, 
leaving the infrared dynamics largely unaffected. 
On the other hand, this ambiguity affects the 
ultraviolet  properties of the SDEs~\cite{Salam:1963sa,Salam:1964zk,Delbourgo:1977jc,Delbourgo:1977hq}.
Essentially, failing  to provide  the correct
transverse  part  leads  to  the mishandling of  overlapping  divergences,
which,  in turn, compromises  the multiplicative  renormalizability of
the resulting SD equations.
The construction of the appropriate transverse part is technically 
complicated, even for QED~\cite{King:1982mk,Curtis:1990zs,Kizilersu:2009kg,Bashir:1997qt},
and its systematic generalization to QCD is still pending 
(for an early attempt in this direction, see \cite{Haeri:1988af}).

An  additional  important  point, not
addressed here,  is related to the  way the $BQQ$  vertex triggers the
Schwinger   mechanism~\cite{Schwinger:1962tn,Schwinger:1962tp},  which,  in   turn,  is
responsible  for the  dynamical  generation  of a  gluon  mass. As  is
well-known~\cite{Jackiw:1973tr,Cornwall:1973ts,Eichten:1974et},  the  relevant   three-gluon  vertex
($\bqq$  in this  case) must  contain longitudinally  coupled massless
poles (last diagram in Fig.~\ref{SDE-vertex}),  
in order  for gauge  invariance  to be  preserved. 
The  Ansatz presented here does not incorporate such poles, 
which must be supplied at a subsequent  step; after this has been  accomplished, 
one can solve numerically the  resulting SDE, and compare with  the available lattice
results.  Work in this direction is currently underway, and we hope to
report on the results in the near future.

\acknowledgments 

The research of J.~P. is supported by the European FEDER and  Spanish MICINN under 
grant FPA2008-02878, and the Fundaci\'on General of the UV.

\end{document}